\documentclass[prd,aps,twocolumn,amsmath,amssymb,nofootinbib,preprintnumbers]
{revtex4}

\usepackage{graphicx}% Include figure files
\usepackage{dcolumn}% Align table columns on decimal point
\usepackage{bm}% bold math
\usepackage{amsmath}
\usepackage{amsfonts}

\newcommand{\remove}[1]{}

\def\ls{\mathrel{\lower4pt\vbox{\lineskip=0pt\baselineskip=0pt
           \hbox{$<$}\hbox{$\sim$}}}}
\def\gs{\mathrel{\lower4pt\vbox{\lineskip=0pt\baselineskip=0pt
           \hbox{$>$}\hbox{$\sim$}}}}
\def\drawbox#1#2{\hrule height#2pt

\hbox{\vrule width#2pt height#1pt \kern#1pt
              \vrule width#2pt}
              \hrule height#2pt}

\def\Asym#1#2{\vcenter{\vbox{\drawbox{#1}{#2}
              \kern-#2pt       % line up boxes
              \drawbox{#1}{#2}}}}

\def\gBL{g_{\scriptscriptstyle B-L}}
\def\QBL{Q_{\scriptscriptstyle B-L}}

\newcommand{\beq}{\begin{equation}}
\newcommand{\eeq}{\end{equation}}

\begin{document}

%
%\vspace*{2cm}
\title{Sneutrino Dark Matter and the Observed Anomalies in Cosmic Rays}

\author{Rouzbeh Allahverdi$^{1}$}
\author{Bhaskar Dutta$^{2}$}
\author{Katherine Richardson-McDaniel$^{1}$}
\author{Yudi Santoso$^{3}$}

\affiliation{$^{1}$~Department of Physics \& Astronomy, University of New Mexico, Albuquerque, NM 87131, USA \\
$^{2}$~Department of Physics, Texas A\&M University, College Station, TX 77843-4242, USA \\
$^{3}$~Institute for Particle Physics Phenomenology, Department of Physics, University of Durham, Durham DH1 3LE, UK}

%\date{February 17, 2009}

\begin{abstract}
We revisit sneutrino dark matter in light of the recent results from the PAMELA and ATIC experiments. In the $U(1)_{B-L}$ extension of the minimal supersymmetric standard model the right-handed sneutrino is a natural candidate for thermal dark matter. Sneutrino annihilation at the present time can be considerably enhanced due to the exchange of the lightest field in the Higgs sector that breaks $U(1)_{B-L}$. The annihilation mainly produces taus (or muons) by the virtue of $B-L$ charges. A sneutrino mass in the $1-2$~TeV range provides a good fit to the PAMELA data and a reasonable fit to the ATIC data. Within this mass range the sneutrino-nucleon elastic scattering cross section is $10^{-11}-10^{-9}$~pb, which might be probed by upcoming and future direct detection experiments. In addition, if (at least) one of the neutrinos is dominantly a Dirac fermion, the sneutrino can provide a unified picture of dark matter and inflation.
\end{abstract}

MIFTP-09-06, IPPP/09/07, DCPT/09/14 \\ February, 2009
\maketitle
%%%%%%%%%%%%%%%%%%%%%%%%%%%%%%%%%%%%%%%%%%%%%%%%%%%%%%%%%%%%%%%%%%%%%%%%%%%%%%%
\section{Introduction}

Even though the existence of dark matter has been supported by various lines of evidence, the identity of dark matter itself is not yet known. One proposed solution for this dark matter problem comes from particle physics beyond the standard model in the form of weakly interacting massive particles (WIMPs)~\cite{Silk}. In particular, for weak scale masses and interactions, thermal freeze out of WIMP annihilation in the early universe can result in an acceptable relic abundance for dark matter, as precisely measured by cosmic microwave background (CMB) experiments~\cite{WMAP5}. Supersymmetry, as one candidate for physics beyond the standard model, has a natural dark matter candidate in the lightest supersymmetric particle (LSP). It is known that in supersymmetric models a neutralino LSP is a suitable candidate for dark matter~\cite{EHNOS}.

There are currently major experimental efforts for both direct and indirect detection of the dark matter particle.
Direct detection probes the scattering of the dark matter particle off nuclei in underground dark matter detectors, while indirect detection investigates astrophysical effects of dark matter annihilation in the galaxy, including signatures in the cosmic rays.
PAMELA is a satellite-borne experiment that measures cosmic ray fluxes. The recently published results show an excess of positron flux at energies above 10~GeV~\cite{PAMELA}, while no excess of anti-proton flux is observed~\cite{PAMELA-antiproton}. The publication shows results up to $\sim 100$~GeV and the experiment is expected to get data up to $\sim 190$~GeV for anti-protons and $\sim 270$~GeV for positrons. Another cosmic ray experiment called ATIC (a balloon experiment) has also recently published data where one observes an excess in the $e^{+} + e^{-}$ spectrum with a peak around 600~GeV~\cite{ATIC}.
There is a third experiment, the PPB-BETS~\cite{pep} (also a balloon experiment), which reports an excess in the $e^+ + e^-$ energy spectrum between 500 and 800 GeV. However, the excess is based on a few data-points that are not quite consistent with the ATIC data. While there could be astrophysical explanations for these anomalies (e.g. from nearby pulsars~\cite{pulsars}), it is reasonable to ask whether they can be attributed to the effect of dark matter annihilation in the galaxy.

Model-independent analysis shows that the annihilation cross section required to explain the positron excess exceeds the canonical value required by relic density, i.e. $\sim 3 \times 10^{-26}$~cm$^3$/s, by at least an order of magnitude~\cite{Vernon}. In the usual neutralino dark matter scenario in the minimal Supergravity (mSUGRA) model, the situation is further complicated because the dark matter annihilation (to fermions) in that model is $P$-wave suppressed, implying a much smaller annihilation cross section today as compared to that at the freeze out time.
An astrophysical boost factor of $10^3-10^4$ is then needed to explain the observed positron excess~\cite{Lars}. However, this might be difficult to obtain based on  recent analyses of halo substructures (see e.g.~\cite{halostructure}). Moreover, in order to explain both the positron and anti-proton data, dark matter annihilation must be dominated by leptonic final state modes~\cite{Strumia,Salati}. (There could also be some effects from anisotropic propagation on the positron and anti-proton fluxes that still need to be investigated~\cite{anisotropic}.)
There have been proposals~\cite{Strumia,Nima} (also see~\cite{Other}) for new dark matter models in which the dark matter candidate belongs to a hidden sector, and  an acceptable thermal relic density is obtained via new gauge interactions. 
The key ideas of these models are that the dark matter annihilation today is enhanced by a Sommerfeld effect~\cite{Sommerfeld} due to the existence of light bosons and that annihilation mainly produces lepton final states via symmetry of the hidden sector. This arrangement explains PAMELA data for a dark matter mass of a few hundred GeV, without needing a large astrophysical boost factor, and ATIC data for larger values of dark matter mass~\cite{Weiner}. Another type of explanation that has been proposed for the data is decaying dark matter with a tuned lifetime~\cite{decaying}.

We recently proposed an explicit model that can explain the measured anomalies in the cosmic rays~\cite{ADRS}. It is based on a simple extension of the minimal supersymmetric standard model (MSSM) that includes a gauged $U(1)_{B-L}$ and where the dark matter is the lightest neutralino in the new sector.
Even though this model has a large dark matter annihilation cross section today due to Sommerfeld enhancement, the cross section for scattering of dark matter off quarks is too low to be accessible to direct detection experiments. In fact, this is a generic situation for hidden sector dark matter models that can explain PAMELA data along the line discussed above.

In this paper we again consider a $B-L$ extension of the MSSM, but with the right-handed (RH) sneutrino as the dark matter. As we will argue, this is a minimal model of thermal dark matter that can explain the observed anomalies in the cosmic rays and can also be probed by  direct detection experiments.
The main channel of sneutrino annihilation is to light Higgs fields, which carry a non-zero $B-L$ quantum number. These Higgs particles in turn decay dominantly to leptons by virtue of the $B-L$ charges for fermions. The same Higgs field also results in a large Sommerfeld enhancement factor for the annihilation cross section. For a sneutrino mass of 1-2 TeV, this model can explain PAMELA and ATIC data. In addition, due to the scalar nature of dark matter, the sneutrino-proton elastic scattering cross section is in the $10^{-11}-10^{-9}$~pb range, which is an interesting range from a direct detection perspective.
Moreover, the sneutrino can be part of the field that drives primordial inflation, thus explaining the small temperature anisotropy in the cosmic microwave background (CMB) via tiny neutrino masses~\cite{AKM,ADM}. We will also discuss various possibilities for radiative breaking of the $B-L$ symmetry and some related issues.

%%%%%%%%%%%%%%%%%%%%
\section{The model}

The $B-L$ extension of the MSSM~\cite{mohapatra} is well motivated since it automatically implies the existence of three RH neutrinos through which one can explain the neutrino masses and mixings. The minimal model contains a new gauge boson $Z^{\prime}$, two new Higgs fields $H^{\prime}_1$ and $H^{\prime}_2$, the RH neutrinos $N$, and their supersymmetric partners. The superpotential is (the boldface characters denote superfields)
\begin{equation} \label{sup}
W = W_{\rm MSSM} + W_{B-L} + y_D {\bf N}^c {\bf H_u} {\bf L} \, ,
\end{equation}
where ${\bf H_u}$ and ${\bf L}$ are the superfields containing the Higgs field that gives mass to up-type quarks and the left-handed (LH) leptons respectively. For simplicity, we have omitted the family indices. The $W_{B-L}$ term contains ${\bf H^{\prime}_1},~{\bf H^{\prime}_2}$ and ${\bf N}^c$. Its detailed form depends on the charge assignments of the new Higgs fields (explicit examples will be presented later). The last term on the RH side of Eq.~(\ref{sup}) is the neutrino Yukawa coupling term.

The scalar potential consists of $F$-terms from the superpotential, and $D$-terms from the gauge symmetries.
The $D$-term contribution from $U(1)_{B-L}$ is given by
\begin{equation} \label{dterm1}
V_D \supset \frac{1}{2} D^2_{B-L} ,
\end{equation}
where
\begin{equation} \label{dterm2}
D_{B-L} = \frac{1}{2} \gBL \left[Q_{1} ({\vert H^{\prime}_1 \vert}^2 - {\vert H^{\prime}_2 \vert}^2) + \frac{1}{2} {\vert \tilde N \vert}^2 + ... \right] .
\end{equation}
Here $\gBL$ is the gauge coupling of $U(1)_{B-L}$, and $+Q_1$, $-Q_1$, $1/2$ are the $B-L$ charges of $H^{\prime}_1,~H^{\prime}_2,~{\tilde N}$ respectively (${\tilde N}$ is the sneutrino field). The $U(1)_{B-L}$ is broken by the vacuum expectation value (VEV) of $H^{\prime}_1$ and $H^{\prime}_2$, which we denote by $v^{\prime}_1$ and $v^{\prime}_2$ respectively. This results in a mass $m_{Z^\prime} = \gBL Q_1 \sqrt{v^{\prime 2}_1 + v^{\prime 2}_2}$ for the $Z^{\prime}$ gauge boson. We have three physical Higgs fields $\phi,~\Phi$ (scalars) and ${\cal A}$ (a pseudo scalar). The scalar Higgses are related to the real parts of $H^{\prime}_1,~H^{\prime}_2$ through the mixing angle $\alpha^{\prime}$:
\begin{eqnarray} \label{vev}
H^{\prime}_1 & = & \frac{v^{\prime}_1 + \cos \alpha^{\prime} \Phi - \sin \alpha^{\prime} \phi}{\sqrt{2}} + \frac{H^{\prime}_{1,I}}{\sqrt{2}}  \, \nonumber \\
H^{\prime}_2 & = & \frac{v^{\prime}_2 + \sin \alpha^{\prime} \Phi + \cos \alpha^{\prime} \phi}{\sqrt{2}} + \frac{H^{\prime}_{2,I}}{\sqrt{2}}  \, ,
\end{eqnarray}
where $H^{\prime}_{1,I}, H^{\prime}_{2,I}$ represent the imaginary parts.
Eqs.~(\ref{dterm1},\ref{dterm2},\ref{vev}) lead to the following terms in the scalar potential
\begin{eqnarray} \label{Ncoupling}
V &\supset& -\frac{1}{2} \gBL m_{Z^\prime} \sin (\alpha^{\prime} + \beta^{\prime}) \, \phi \, {\vert \tilde N \vert}^2 \, \nonumber \\
&& - \frac{1}{2} \gBL^{2} Q_1 \cos (2 \alpha^{\prime}) \, \phi^2 \, {\vert \tilde N \vert}^2 \, \nonumber \\
&& + \frac{1}{2} \gBL m_{Z^\prime} \cos (\alpha^{\prime} + \beta^{\prime}) \, \Phi \, {\vert \tilde N \vert}^2 \, \nonumber \\
&& + \ ... \ ,
\end{eqnarray}
where $\tan \beta^{\prime} \equiv v^{\prime}_2/v^{\prime}_1$. The masses of the Higgs fields follow $m^2_{\phi} < \cos^2 (2 \beta^{\prime}) \, m^2_{Z^\prime}$ and $m_{\Phi},~m_{\cal A} \sim m_{Z^\prime}$.

A natural dark matter candidate in this model is the sneutrino ${\tilde N}$~\footnote{Another candidate is the lightest neutralino in the new sector, which is a linear combination of the $U(1)_{B-L}$ gaugino ${\widetilde Z}^{\prime}$ and the two Higgsinos ${\widetilde H}^{\prime}_1$, ${\widetilde H}^{\prime}_2$~\cite{ADRS,khalil}.}. We note that the ${\tilde N}$ has fewer gauge interactions than other fields, hence its mass receives the smallest contribution from the gaugino loops. The main processes for annihilation of dark matter quanta are then governed by interactions in Eq.~(\ref{Ncoupling}). The dominant channel is ${\tilde N}^* {\tilde N} \rightarrow \phi \phi$ via the $s$-channel exchange of the $\phi,~\Phi$, the $t$,~$u$-channel exchange of the ${\tilde N}$, and the contact term $\vert {\tilde N} \vert^2 \phi^2$. The $s$-channel $Z^\prime$ exchange is subdominant because of the large $Z^\prime$ mass (as required by the experimental bound on $m_{Z^\prime}$).
There are also ${\tilde N}^* {\tilde N} \rightarrow \phi \Phi , ~ \phi {\cal A}, ~ \Phi \Phi,~{\cal A} {\cal A}$ annihilation processes, but they are kinematically suppressed and/or forbidden for the parameter space we are considering. The sneutrinos can also annihilate to RH neutrinos via $t$-channel neutralino exchange. Again for the parameter space that we consider the annihilation into $\nu \bar{\nu}$ final states is at least an order of magnitude below the $\phi \phi$ final states. Other fermion final states, through $s$-channel $Z^\prime$ exchange, have even smaller branching ratios. Moreover, note that the annihilations to fermion-antifermion final states are $P$-wave suppressed.

The $\phi$ subsequently decays into fermion-antifermion pairs via a one-loop diagram containing two $Z^{\prime}$ bosons. The decay rate is given by:
\begin{equation} \label{phidec}
\Gamma(\phi \to f {\bar f}) = \frac{C_f}{2^7 \pi^5} \frac{\gBL^6 Q^4_{f} Q^2_{\phi} m^5_{\phi} m^2_f}{m^6_{Z^{\prime}}} \left(1 - \frac{4 m^2_f}{m^2_{\phi}} \right)^{3/2},
\end{equation}
where $Q_f$ and $Q_\phi$ are the $B-L$ charges of the final state fermion and the $\phi$ respectively, $m_f$ is the fermion mass, and $C_f$ denotes color factor.
Since the $B-L$ charge of leptons is three times larger than that of quarks, the leptonic branching ratio is naturally larger than that for quarks. We note that $m_{\phi}$ can be controlled by the VEVs of the new Higgs fields and for comparable VEVs, i.e. for $\tan \beta^{\prime} \approx 1$, it can be very small compared to $m_{Z^\prime}$. For $m_{\phi} > 2 \, m_b$ the dominant decay mode is $\phi \rightarrow \tau^{-} \tau^{+}$ final state, while the branching ratio for the $\phi \rightarrow b {\bar b}$ mode is $\approx 7$ times smaller.

The annihilation cross section at the present time has Sommerfeld enhancement as a result of the attractive force between sneutrinos due to the $\phi$ exchange. The Higgs coupling to dark matter is given by the first term on the RH side of Eq.~(\ref{Ncoupling}) and leads to an attractive potential $V(r) = -\alpha (e^{-m_{\phi}r}/r)$ in the non-relativistic limit~\cite{Sommerfeld}, where
\begin{equation} \label{fsc}
\alpha = \frac{\gBL m_{Z^\prime} \sin (\alpha^{\prime} + \beta^{\prime})}{4 m_{\tilde N}} \ ,
\end{equation}
and $m_{\tilde N}$ is the sneutrino mass.

%%%%%%%%%%%%%%%%%%%%%%%%%%%%%%%%%%
\section{Sneutrino dark matter and PAMELA}

\begin{figure}[ht]
%\vspace*{1.0cm}
\begin{center}
%\vskip -0.8in
\includegraphics[width=.48\textwidth]{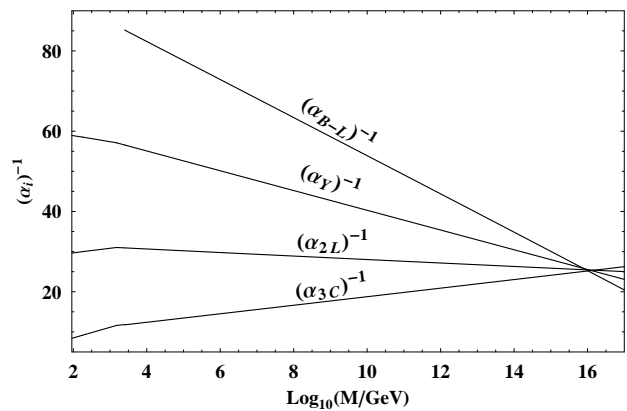}
\end{center}
%\vskip -0.65in
\caption{We show the unification of gauge couplings for the $B-L$ charge assignments in Table 1 using two loop RGEs.}
\label{unification}
\end{figure}

We are now going to show that the sneutrino dark matter can explain the PAMELA data.
We first identify the allowed regions of the model parameter space that result in an acceptable dark matter relic density and then find the Sommerfeld enhancement factor for these regions. For an explicit example, which we call the minimal model, we choose the $B-L$ charge for $H_1^\prime$ (i.e. $Q_1$) to be $3/2$.
The $B-L$ charges of the fields involved are shown in Table~\ref{BLcharges-tbl}.
\begin{table}[tbp]
\center
\begin{tabular}{|c||c|c|c|c|c|c|c|}\hline
{\rm Fields} & $Q$ & $Q^c$ & $L$ & $L^c$ &  $H^{\prime}_1$ & $H^{\prime}_2$ \\ \hline
$\QBL$ & 1/6 & -1/6 & -1/2 & 1/2 &  3/2 & -3/2 \\ \hline
\end{tabular}
\caption{The $B-L$ charges of the fields for the minimal model. Here $Q$ and $L$ represent quarks and leptons respectively, while $H^{\prime}_1$ and $H^{\prime}_2$ are the two new Higgs fields. The MSSM Higgs fields have zero $B-L$ charges.}
\label{BLcharges-tbl}
\end{table}
We use reasonable values for the model parameters, i.e., $\tan \beta^\prime \approx 1$, $m_{Z^\prime} > 1.5$~TeV, $\mu^\prime = 0.5 - 1.5$~TeV ($\mu^{\prime}$ being the Higgs mixing parameter in the $B-L$ sector), soft masses for the Higgs fields $m_{H_{1,2}^\prime} = 200-600$~GeV, and soft gaugino mass $M_{\widetilde{Z}^\prime} \geq 1$~TeV. We use $\gBL \sim 0.40$, which is in concordance with unification of the gauge couplings (we need to use a normalization factor $\sqrt{3/2}$ for unification).
We show the unification of all the gauge couplings using the two loop renormalization group equations (RGEs) in Figure~\ref{unification}. We find that for $m_{Z^{\prime}} \simeq 2.5$ TeV the couplings unify at $\sim 10^{16}$ GeV. This figure is drawn for the $B-L$ assignments shown in Table~\ref{BLcharges-tbl}.
The $Z^{\prime}$ mass used in the calculation obeys the LEP and Tevatron bounds~\cite{tev,carena} for our charge assignments. The sneutrino mass is chosen to be between 800 GeV and 2 TeV in order to explain the PAMELA (and ATIC) data.

\begin{figure}[ht]
%\vspace*{0.5cm}
\begin{center}
\includegraphics[width=8.0cm]{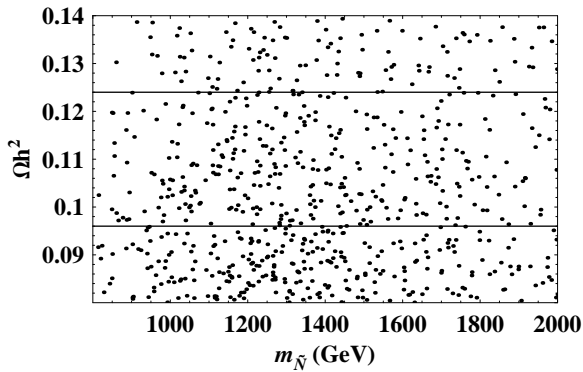}
\end{center}
\vskip -0.25in
\caption{We show the relic density and the sneutrino mass for model points generated by varying the parameters mentioned in the text. \label{ohsqvsm}}
\label{model}
\end{figure}

\begin{figure}[ht]
%\vspace*{0.5cm}
\begin{center}
\includegraphics[width=8.0cm]{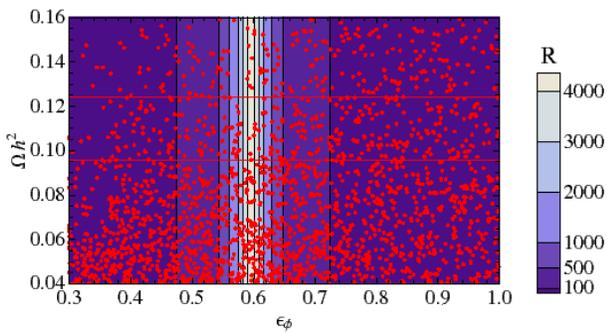}
\end{center}
\vskip -0.25in
\caption{We show relic density as a function of $\epsilon_\phi$. We show different ranges for the Sommerfeld enhancement factor $R$ by shaded contours.\label{ohsqvsephi}}
\end{figure}

In Figure~\ref{ohsqvsm}, we show the relic density and the sneutrino mass for different model points. The horizontal band shows the acceptable range for relic density according to the latest WMAP data~\cite{WMAP5}. In Figure~\ref{ohsqvsephi} we show the possible Sommerfeld enhancement factor $R$ that can be obtained for these points in term of $\epsilon_\phi \equiv m_{\phi}/\alpha m_{\tilde N}$. Note that many of the points that satisfy the relic density constraint have enhancement factor $R \geq 10^3$, corresponding to $\epsilon_\phi = 0.55$ to $0.65$. This is true for the whole
range of $m_{\tilde N}$ shown in Figure~\ref{ohsqvsm}. The lifetime of $\phi$ for these points is found to be $\tau_{\phi} \sim 10^{-5}-10^{-4}$~seconds from Eq.~(\ref{phidec}). Thus $\phi$'s produced in the early universe decay rapidly enough in order not to affect big bang nucleosynthesis (BBN)~\cite{BBN}\footnote{Since dark matter particles are non-relativistic at the time of BBN, their annihilation enhanced by Sommerfeld effect can result in significant electromagnetic and/or hadronic showers that dissociate light elements from BBN. For muon final states, the large enhancement factor required to explain PAMELA is compatible with BBN bounds, while for tau final states a small astrophysical boost factor seems to be needed in order not to affect BBN~\cite{BBN2}.}.
We select points that satisfy the dark matter relic density and then use {\tt DarkSUSY}-5.0.2~\cite{darksusy} to calculate the positron flux from dark matter annihilation. Each pair annihilation in our model produces 2 $\phi$'s that yield four fermions upon their decay. For this reason, we generally need a heavier sneutrino compared to models in which the pair annihilation directly produces fermions. We normalize the positron fraction by a factor $k_b = 1.11$~\cite{BE}. There are theoretical uncertainties in the positron cosmic ray flux calculation due to the assumptions about the dark matter halo profile and the cosmic ray propagation model. Here we assume NFW profile~\cite{NFW} for the dark matter halo and MED parameters for the propagation as defined in~\cite{Delahaye}.

\begin{figure}[ht]
%\vspace*{1.0cm}
\begin{center}
\vskip -0.9in
\includegraphics[width=.48\textwidth]{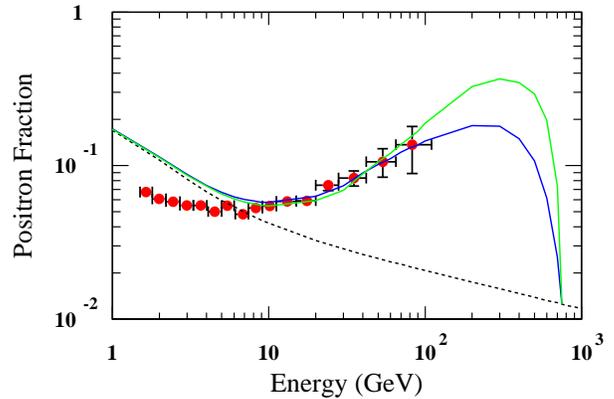}
\end{center}
\vskip -0.65in
\caption{We show a fit to the PAMELA data when the $\phi$ decays mostly to taus (dark blue) or muons (light green) for a sneutrino mass of 1.5~TeV and an enhancement factor of $10^3$. The dashed line is the expected background cosmic rays.}
\label{pamelatau}
\end{figure}

In Figure~\ref{pamelatau}, we show our fit to the PAMELA data for $m_{\tilde N} = 1.5$~TeV for $\tau^+ \tau^-$ and $\mu^+ \mu^-$ final state cases. We found that with an enhancement factor of $10^3$ the chi-square values (including only points with energy greater than 10~GeV) for a sneutrino mass of 1.5~TeV are small, i.e. 2.9 and 5.5 for $\tau^+ \tau^-$ and $\mu^+ \mu^-$ respectively. When $m_{\phi}$ is (chosen to be) below $2 \, m_b$ but above $2 \, m_{\tau}$, we do not have any anti-proton excess. In fact we can raise $m_{\phi}$ up to $\sim15$~GeV and still have acceptable anti-proton flux.
We can also have a reasonable fit to the ATIC data, although simultaneous fit for both ATIC and PAMELA are not satisfactory~\cite{ADRS}.

We note that this model also has a great potential to be observed with the Fermi Satellite experiment. Due to electromagnetically charged final states of $\phi$ decays, the Sommerfeld enhancement would also lead to a higher rate of photons in the gamma ray background~\cite{CRgamma}. There could also be contribution to the neutrino flux~\cite{CRneutrino}.

%%%%%%%%%%%%%%%%%%%%%%%%%%%%%%%%%%
\section{Direct Detection}

\begin{figure}[ht]
%\vspace*{1.0cm}
\begin{center}
%\vskip -0.8in
\includegraphics[width=.48\textwidth]{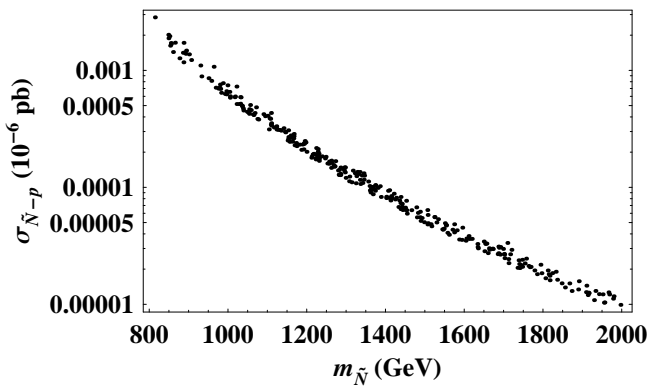}
\end{center}
\vskip -0.25in
\caption{We show the direct detection cross section as a function of sneutrino mass.}
\label{direct}
\end{figure}

The current upper bound on the spin-independent dark matter particle-proton scattering cross section is about $4.6 \times 10^{-8}$~pb for a dark matter mass around 60~GeV, and increasing to $\sim 2 \times 10^{-7}$~pb for a mass around 1.2~TeV~\cite{cdms}. In our model the elastic scattering of the sneutrino occurs via the $Z^{\prime}$ exchange with the nucleus in the $t$-channel. This leads to only a spin-independent contribution since the $B-L$ charges of the left and right quarks are the same. In Figure~\ref{direct}, we show the $\tilde N$-$p$ scattering cross section for the model points in Figure 1 that satisfy the relic density constraint $0.096 < \Omega_{DM} h^2 < 0.124$. We see that the cross section can be in the $10^{-11}-10^{-9}$ pb range,
which is close to the reach of
the upcoming dark matter direct detection experiments~\cite{Direct}.

It is also seen that the cross section decreases as the sneutrino mass increases. This is because for larger values of $m_{\tilde N}$ we also need a larger annihilation cross section to satisfy the relic density constraint. As discussed earlier the annihilation cross section depends on the sneutrino couplings to $\phi$ and $\Phi$, see Eq.~(\ref{Ncoupling}), which are $\propto m_{Z^\prime}$.

It is interesting to note that within this mass range the $Z^{\prime}$ can be produced at the LHC. The $Z^{\prime}$ decay will produce the new light Higgs $\phi$ (among other fields). However, $\phi$ will decay outside of the detector because of its relatively long life time ($\sim 10^{-5}$ sec).
Thus, in addition to the sneutrino LSP, we have another source of missing energy signal in this model~\footnote{We also note that there are 7 neutralinos in this model, compared to four in the MSSM, while the number of charginos is still two. Therefore, using the end point analysis~\cite{endpoint}, one can find many neutral states.}.

%%%%%%%%%%%%%%%%%%%%%%%%%%%%%%%%%%
\section{Radiative breaking of $B-L$ symmetry}

In order to have spontaneous breaking of $U(1)_{B-L}$ we need a negative eigenvalue in the $H^{\prime}_1,~H^{\prime}_2$ square-mass matrix. This can arise dynamically as a result of radiative corrections to the Higgs soft masses. Here we discuss two generic possibilities.

%%%%%%%%%%%%%%%%%%%%%%%%
\subsection{Higgs coupling to right-handed neutrinos}

If $Q_1 = 1$, then $H^{\prime}_2$ can have a superpotential coupling to the RH neutrinos~\footnote{Similarly, $H^{\prime}_1$ can couple to the RH neutrinos if $Q_1 = -1$.}. In this case we have:
\begin{equation} \label{sup1}
W_{B-L} = f {\bf H^{\prime}_2} {\bf N}^c {\bf N}^c  + \mu^{\prime} {\bf H^{\prime}_1} {\bf H^{\prime}_2} ,
\end{equation}
where $\mu^{\prime}$ is analogous to the MSSM $\mu$ parameter. Note that in the minimal model, with $Q_1 = 3/2$, the $H^{\prime}_2$ can only couple to $H^{\prime}_1$ in the superpotential. Taking the soft mass parameters into account, the Higgs potential is
\begin{equation} \label{soft}
(m^2_1 + \mu^{\prime 2}) {\vert H^{\prime}_1 \vert}^2 + (m^2_2 + \mu^{\prime 2}) {\vert H^{\prime}_2 \vert}^2 + \left(B \mu^{\prime} H^{\prime}_1 H^{\prime}_2 + {\rm h.c} \right).
\end{equation}
(Here the parameters $m_1, m_2, B$ are not to be confused with the MSSM Higgs parameters.)
The Yukawa coupling between ${\bf H^{\prime}_2}$ and ${\bf N}^c$ can drive $m^2_2$ to a sufficiently negative value such that $m^2_2 + \mu^{\prime 2} < 0$ around the TeV energy scale~\footnote{Since $H^{\prime}_1$ has no Yukawa couplings, $m^2_1$ increases towards smaller scales because of the $U(1)_{B-L}$ gaugino loop.}. This requires that $f$ not be very small. On the other hand, $f$ should not be very large. Otherwise there would be a one-loop correction that lifts the $\phi$ mass above its tree-level limit $m^2_{\phi} < M^2_{Z^{\prime}} {\rm cos}^2(2 \beta^{\prime})$ (similar to the correction from the top Yukawa in the MSSM~\cite{AN}):
\begin{equation} \label{lifting}
\Delta m^2_{\phi} \sim \frac{1}{16 \pi^2} f^2 m^2_N {\rm ln} \left[\frac{m^2_N}{m_{{\tilde N}_R} m_{{\tilde N}_I}} \right].
\end{equation}
Here $m_N = \sqrt{2} f v^{\prime}_2$ is the Majorana mass of $N$ and
\begin{eqnarray} \label{split}
m_{{\tilde N}_R} & = & \left[m^2_{\tilde N} + m_N (A + \mu^{\prime}) \right]^{1/2} \, \nonumber \\
m_{{\tilde N}_I} & = & \left[m^2_{\tilde N} - m_N (A + \mu^{\prime}) \right]^{1/2} \, ,
\end{eqnarray}
are the total masses of the real and imaginary parts of the ${\tilde N}$ respectively. The mass splitting is due to the contribution of the $A$ and $\mu^{\prime}$ terms to the ${\tilde N}$ potential given by
\begin{equation} \label{sneutrino}
f \left( A H^{\prime}_2 + \mu^{\prime} H^{\prime *}_1 \right) {\tilde N} {\tilde N} + {\rm h.c} .
\end{equation}
For simplicity we have assumed that $f,~A$ and $~\mu^{\prime}$ are real, and we have used the requirement that $v^{\prime}_1 \approx v^{\prime}_2$. This mass splitting can lift $m_\phi$ above its tree-level limit similar to the effect from stop mixing in the MSSM~\cite{AN}. In order for this not to be significant, it will be sufficient to have $\vert A + \mu^{\prime} \vert \ll m_{\tilde N}$ at the TeV scale~\footnote{This also ensures that the contribution of Eq.~(\ref{sneutrino}) to the coupling of dark matter to $\phi,~\Phi$ is negligible compared with that from the $D$-term~(\ref{Ncoupling}).}.

In our case the sneutrino  and the neutrino masses can be close and therefore the corrections to the Higgs mass can be very small $<10$ GeV (which we have checked numerically). In fact, this allows us to keep the Higgs mass below $15$ GeV such that we do not have any problems with anti-proton data. We also have found that $f \sim 0.2$ is large enough to lead to radiative breaking of $U(1)_{B-L}$. Assuming grand unification, one should use the SUGRA boundary conditions to achieve radiative breaking due to the smallness of $f$ (as compared with the top Yukawa in the MSSM). The appropriate boundary conditions are such that $m^2_1 = m^2_2 < 0$, but $m^2_{1,2} + \mu^{\prime 2} > 0$, while all other soft $({\rm mass})^2$ are positive at the grand unification scale.

The potential term in Eq.~(\ref{sneutrino}), no matter how small, inevitably splits the masses of the real and imaginary parts of the sneutrino. In this case dark matter will be the lighter of ${\tilde N}_R$ and ${\tilde N}_I$, which we choose to be ${\tilde N}_R$ without loss of generality. Then the heavier component ${\tilde N}_I$ decays to dark matter and fermions via an off-shell $Z^{\prime}$ in the early universe. In this case the dark matter coupling to the light Higgs $\phi$, the first term on the RH side of Eq.~(\ref{Ncoupling}), will be twice as small as the term wherein the dark matter is a complex field. However the attractive potential, due to $\phi$ exchange, will be the same in the two cases. This is because a real field annihilates and creates the same quanta, thus giving rise to a factor of two that compensates the $1/2$ that appears in the coupling. The same also holds for dark matter annihilation. Therefore, as far as relic density calculations and the Sommerfeld enhancement factor are concerned, our results will not depend on whether dark matter is the complex ${\tilde N}$ field or its real (or imaginary) component.

We also note that the $f {\bf H^{\prime}_2} {\bf N}^c {\bf N}^c$ term opens a new channel for the decay of the light Higgs $\phi$. Since $m_{\phi} \ll m_N$, the decay to on-shell RH neutrinos is not possible. The decay can happen via off-shell $N$ through their couplings to the MSSM fields~(\ref{sup}), at higher orders of perturbation theory. Then, if the neutrino Dirac Yukawa coupling $y_D$ is sufficiently small, this decay mode will be totally negligible. Indeed this is the case since the see-saw mechanism at the TeV scale requires $y_D \ls 10^{-6}$ to generate the observed neutrino masses.

%%%%%%%%%%%%%%%%%%%%%%%%%%%%%%%%%%%
\subsection{Higgs coupling to other fields}

If $Q_1 \neq \pm 1$, then neither $H^{\prime}_1$ nor $H^{\prime}_2$ can have a superpotential coupling to $N$. This happens, for example, if $Q_1 = 1/2$~\cite{ADM} or $Q_1 = 3/2$ (as chosen above for the minimal model and also in~\cite{ADRS}). Radiative breaking of $U(1)_{B-L}$ then requires a moderate Yukawa coupling of one of the $B-L$ Higgses to the new fields.

A simple model of this type includes two new superfields ${\bf \Phi_1},~{\bf \Phi_2}$ and has the following superpotential
\begin{eqnarray} \label{sup2}
W_{B-L} & = & \mu^{\prime} {\bf H^{\prime}_1} {\bf H^{\prime}_2} + f_1 {\bf H^{\prime}_1} {\bf \Phi_1} {\bf \Phi_1} + f_2 {\bf H^{\prime}_2} {\bf \Phi_2} {\bf \Phi_2} \, \nonumber \\
& + & \mu^{\prime \prime} {\bf \Phi_1} {\bf \Phi_2} +  \lambda {\bf \Phi_1} {\bf N}^c {\bf N}^c \, , \nonumber \\
&& \,
\end{eqnarray}
where $Q_1 = 2$ and the $B-L$ charges of ${\bf \Phi_1},~{\bf \Phi_2}$ are $-1,~+1$ respectively. All we need for radiative breaking of $U(1)_{B-L}$ is $f_1 \sim 0.2$, or $f_2 \sim 0.2$ (one of the couplings can be very small) and SUGRA boundary conditions. This ensures sufficiently large loop corrections that drive the $({\rm mass})^2$ of the corresponding Higgs field to negative values around a TeV. In this case, since the Higgs fields are not coupled to ${\bf N}^c$, there are no Majorana masses for the RH neutrinos, and hence neutrinos are of a Dirac nature. Also, there will be no splitting between the masses of the real and imaginary parts of the ${\tilde N}$, and thus the dark matter is a complex scalar field.

Interactions in the first line of Eq.~(\ref{sup2}) result in a $2 \times 2$ mass matrix for ${\bf \Phi_1},~{\bf \Phi_2}$. For large enough $\mu^{\prime \prime}$ (i.e. $\mu^{\prime \prime} \sim v^{\prime}_1,~v^{\prime}_2$) the mass eigenvalues will be larger than the sneutrino mass. Hence the corresponding mass eigenstates would quickly decay to the RH neutrino and sneutrino through the $\lambda {\bf \Phi_1} {\bf N}^c {\bf N}^c$ superpotential term in the early universe. The $f_1 {\bf H^{\prime}_1} {\bf \Phi_1} {\bf \Phi_1}$ and $\lambda {\bf \Phi_1} {\bf N}^c {\bf N}^c$ terms also open a new decay channel for the light Higgs: $\phi \rightarrow 4N$ via off-shell fermionic components of $\Phi_1$~{\footnote{Note that there can be no ${\tilde N}$ in the final state since $m_{\phi} \ll m_{\tilde N}$.}. However, $\lambda$ can be chosen to be sufficiently small such that $\phi \rightarrow \tau^{-} \tau^{+}$ remains the dominant decay mode of $\phi$.

If $Q_1 = 3/2$, as chosen in Table 1, one can introduce four new superfields ${\bf \Phi_1},~{\bf \Phi_2},~{\bf \Phi_3},~{\bf \Phi_4}$ (with respective $B-L$ charges $-1,~+1,~-1/2,~+1/2$) and the following superpotential
\begin{eqnarray} \label{sup3}
W_{B-L} & = & \mu^{\prime} {\bf H^{\prime}_1} {\bf H^{\prime}_2} + f_1 {\bf H^{\prime}_1} {\bf \Phi_1} {\bf \Phi_3} +  f_2 {\bf H^{\prime}_2} {\bf \Phi_2} {\bf \Phi_4} \, \nonumber \\
& + & \mu^{\prime \prime} {\bf \Phi_1} {\bf \Phi_2} + \mu^{\prime \prime \prime} {\bf \Phi_3} {\bf \Phi_4} + \lambda {\bf \Phi_1} {\bf N}^c {\bf N}^c \, . \nonumber \\
&& \,
\end{eqnarray}
Again the mass eigenvalues of the $4 \times 4$ mass matrix for ${\bf \Phi_1},~{\bf \Phi_2},~{\bf \Phi_3},~{\bf \Phi_4}$ can be made large enough such that the corresponding eigenstates rapidly decay to lighter fields via the $\lambda {\bf \Phi_1} {\bf N}^c {\bf N}^c$ superpotential terms.

It is interesting to note that although we have used $Q_1 = 3/2$ to obtain the results in the previous sections, they are largely independent from the exact charge assignments of $H^{\prime}_1,~H^{\prime}_2$.
This is because the major contributions to relic density calculations involve $Z^{\prime}$ mass in the coupling and $Q_1$ is already absorbed in the mass definition, see Eq.~(\ref{Ncoupling}). The direct detection cross section also remains unchanged for the same reason. The gauge coupling unification still occurs but requires a larger value of $g_{B-L}$ for $\vert Q_1 \vert < 3/2$. Therefore Eqs.~(\ref{sup1},\ref{sup2},\ref{sup3}) can all yield thermal sneutrino dark matter with a large Sommerfeld enhancement factor and radiative breaking of $U(1)_{B-L}$.

%%%%%%%%%%%%%%%%%%%%%%%%%%%%%%%%%%
\section{Right-handed sneutrino and inflation}

In addition to being the dark matter candidate, the RH sneutrino can also drive inflation in the context of the $U(1)_{B-L}$ model~\cite{AKM,ADM}. The gauge-invariant combination ${\bf N}^c {\bf H_u} {\bf L}$ forms a $D$-flat direction under the whole gauge symmetry $SU(3)_C \times SU(2)_L \times U(1)_Y \times U(1)_{B-L}$. The flat direction field $\varphi$ is defined as
\begin{equation} \label{flat}
{\varphi} = \frac{{\tilde N} + {H}_u + {\tilde L}}{\sqrt{3}} .
\end{equation}
The potential along the flat direction, after the minimization along the angular direction, is found to be~\cite{AKM},
\begin{eqnarray} \label{flatpot}
V (\vert \varphi \vert) = \frac{m^2_{\varphi}}{2} \vert \varphi \vert ^2 + \frac{y^2_D}{12} \vert \varphi \vert^4 - \frac{A y_D}{6\sqrt{3}} \vert \varphi \vert^3,
\end{eqnarray}
where $y_D$ is the neutrino Dirac Yukawa~(\ref{sup}) and $A$ is the corresponding $A$-term coupling. The flat direction mass $m_{\varphi}$ is given in terms of the ${\tilde N},~H_u,~{\tilde L}$ masses:
\begin{equation} \label{phimass}
m^2_{\varphi} = \frac{m^2_{\tilde N} + m^2_{H_u} + m^2_{\tilde L}}{3}.
\end{equation}
For $A \approx 4 m_{\varphi}$, there exists an inflection point $\varphi_0$ for which $V^{\prime \prime}(\varphi_0) = 0$. Due to the extreme flatness of the potential around the inflection point, inflation can take place near $\varphi_0$. The amplitude of density perturbations generated during inflation follows~\cite{AKM}
\begin{equation} \label{amp}
\delta_{H} \simeq 3.5 \times 10^{-27} ~ \left( \frac{y_D \langle H_u \rangle}{0.05 ~ {\rm eV}} \right)^2 ~ \left( \frac{M_{\rm P}}{m_{\varphi}} \right) ~ {\cal N}_{\rm COBE}^2\,,
\end{equation}
where $\langle H_u \rangle \simeq 174$ GeV and ${\cal N}_{\rm COBE} \sim 50$ is the number of e-foldings between the time that relevant perturbations were produced and the end of inflation. It is seen from Eq.~(\ref{amp}) that perturbations of the correct size $\delta_H = 1.91 \times 10^{-5}$ are obtained if $y_D \sim 10^{-12}$. Interestingly this is the typical neutrino Dirac Yukawa coupling that gives rise to the mass scale $m_{\nu} \simeq 0.05$ eV required to explain the atmospheric neutrino oscillations detected by the Super-Kamiokande experiment~\cite{atmos} if neutrinos are dominantly Dirac in nature.

Dirac neutrinos can be achieved in both cases considered above~(\ref{sup1},\ref{sup2}). If the Higgs fields that break $U(1)_{B-L}$ are not coupled to ${\bf N}^c$~(\ref{sup2}), the neutrinos are naturally Dirac since there will be no Majorana masses for the RH neutrinos in this case, regardless of how big the couplings $f_1,~f_2$ are. On the other hand, a superpotential coupling between the Higgs that breaks $U(1)_{B-L}$ and ${\bf N}^c$, see Eq.~(\ref{sup1}), inevitably induces a Majorana mass $m_N = 2f \langle H^{\prime}_2 \rangle$ to the RH neutrinos upon the $B-L$ breaking. Nevertheless, the main contribution to the mass of light neutrinos comes from the Dirac Yukawa coupling $y_D$. Hence all that we need in this case is one of the Majorana masses (out of the three generations) to be very small in order to have an almost Dirac neutrino with $y_D \sim 10^{-12}$. The other Majorana masses (and respectively the coupling $f$) can be large.

We therefore conclude that it is possible to have a unified $U(1)_{B-L}$ model of inflation and dark matter, where the RH sneutrino is the dark matter and a component of the inflaton field~\cite{ADM}.

%%%%%%%%%%%%%%%%%%%%%%%%%%%%%%%%%%%%%%%%%%%%%%%%%%%%%%%%
\section{Conclusion}

Motivated by the recently reported cosmic ray anomalies, we have reconsidered a minimal extension of the MSSM that includes a gauged $U(1)_{B-L}$. This additional symmetry is broken around a TeV by two new Higgs fields that carry non-zero $B-L$ charges. The RH sneutrino can naturally be the dark matter candidate in this model since it has the smallest gauge interactions among all the fields. Sneutrino interactions of gauge strength yield an acceptable thermal relic density in large regions of the parameter space. If the lightest Higgs in the $B-L$ sector is much lighter than a TeV, the dark matter dominantly annihilates into final states including this Higgs. The annihilation is governed by $D$-term couplings between the sneutrino and the Higgs and takes place in the $S$-wave.
The light Higgs subsequently decays to fermions and the $B-L$ symmetry guarantees that the branching ratio for producing leptons is several times larger than that for quarks, which agrees with the observation of positron and anti-proton fluxes by PAMELA. For a $1-2$ TeV sneutrino, a Higgs mass $\leq 15$ GeV will result in a large Sommerfeld enhancement factor ${\cal O}(10^3)$ in the annihilation cross section at the present time. This provides a good fit to the PAMELA data and a reasonable fit to the ATIC data.

The sneutrino interacts with quarks via $t$-channel exchange of the $U(1)_{B-L}$ gauge boson $Z^{\prime}$. The interaction only has a spin-independent part since  $B-L$ is a vector symmetry. The resulting sneutrino-nucleon elastic scattering cross section is found to be $10^{-11}-10^{-9}$~pb, which might be within the reach of future direct detection experiments.

We have also discussed radiative breaking of $U(1)_{B-L}$. This requires that (one of) the $B-L$ Higgs fields  have a relatively large Yukawa coupling to the RH sneutrino or some other field.
A Yukawa coupling $\sim 0.2$ is sufficient to induce radiative breaking while keeping the mass of the light Higgs low enough to give rise to considerable Sommerfeld enhancement of dark matter annihilation.

Finally, if (at least) one of the neutrinos is dominantly a Dirac fermion, the sneutrino can be part of the inflaton field in addition to being the dark matter. This is a very appealing scenario since direct and indirect detection experiments not only probe dark matter in this case, but they can also reveal the interactions of the inflaton, which is supposed to be the most elusive particle in the universe.

%%%%%%%%%%%%%%%%%%%%%%%%%%%%%%%%%%%%%%%%%%%%%%%%%%%%%%%%
\section{Acknowledgement}
The work of BD is supported in part by DOE grant DE-FG02-95ER40917.
%%%%%%%%%%%%%%%%%%%%%%%%%%%%%%%%%%%%%%%%%%%%%%%%%%%%%%%%%


\begin{thebibliography}{99}

\bibitem{Silk}
H. Goldberg, Phys. Rev. Lett. \textbf{50} (1983) 1419.

\bibitem{WMAP5}
E. Komatsu, {\it et al.}, arXiv:0803.0547.
% [astro-ph].

\bibitem{EHNOS}
J.R.~Ellis, J.S.~Hagelin, D.V.~Nanopoulos, K.A.~Olive and M.~Srednicki, Nucl.\ Phys.\  B {\bf 238}, 453 (1984).


\bibitem{PAMELA}
O. Adriani, {\it et al.}, arXiv:0810.4995.
% [astro-ph].

\bibitem{PAMELA-antiproton}
O. Adriani, {\it et al.}, arXiv:0810.4994.
% [astro-ph].

\bibitem{ATIC}
J. Chang, {\it et al.}, Nature 456, 362 (2008).


\bibitem{pep}
S.~Torii, {\it et al.}, arXiv:0809.0760.
% [astro-ph].


\bibitem{pulsars}
D.~Hooper, P.~Blasi and P.D.~Serpico, arXiv:0810.1527;
% [astro-ph].
%\bibitem{Profumo:2008ms}
  S.~Profumo,
  %``Dissecting Pamela (and ATIC) with Occam's Razor: existing, well-known
  %Pulsars naturally account for the 'anomalous' Cosmic-Ray Electron and
  %Positron Data,''
  arXiv:0812.4457.
%[astro-ph].
  %%CITATION = ARXIV:0812.4457;%%


\bibitem{Vernon}
V. Barger, W.-Y. Keung, D. Marfatia and G. Shaughnessy, arXiv:0809.0162.
% [hep-ph].

\bibitem{Lars}
L. Bergstrom, T. Bringmann and J. Edsjo, arXiv:0808.3725.
% [astro-ph].

\bibitem{halostructure}
N.~Afshordi, R.~Mohayaee and E.~Bertschinger, arXiv:0811.1582.
% [astro-ph].

\bibitem{Strumia}
M. Cirelli, M. Kadastik, M. Raidal and A. Strumia; arXiv:0809.2409.
% [hep-ph].

\bibitem{Salati}
F. Donato, D. Maurin, P. Brun, T. Delahaye and P. Salati, arXiv:0810.5292.
% [astro-ph].

\bibitem{anisotropic}
%\bibitem{deBoer:2009rg}
  W.~de Boer,
  %``Indirect Dark Matter Signals from EGRET and PAMELA compared,''
  arXiv:0901.2941. % [hep-ph].
  %%CITATION = ARXIV:0901.2941;%%


\bibitem{Nima}
N.~Arkani-Hamed, D.P.~Finkbeiner, T.~Slatyer and N.~Weiner, arXiv:0810.0713.
% [hep-ph].


\bibitem{Other}
%\bibitem{Grajek:2008jb}
P.~Grajek, G.~Kane, D.J.~Phalen, A.~Pierce and S.~Watson,
%``Neutralino Dark Matter from Indirect Detection Revisited,''
arXiv:0807.1508;
%[hep-ph];
%%CITATION = ARXIV:0807.1508;%%
%\bibitem{Pospelov:2008jd}
M.~Pospelov and A.~Ritz,
%``Astrophysical Signatures of Secluded Dark Matter,''
arXiv:0810.1502;
% [hep-ph];
%%CITATION = ARXIV:0810.1502;%%
%\bibitem{Cholis:2008qq}
I.~Cholis, D.P.~Finkbeiner, L.~Goodenough and N.~Weiner,
%``The PAMELA Positron Excess from Annihilations into a Light Boson,''
arXiv:0810.5344;
% [astro-ph];
%%CITATION = ARXIV:0810.5344;%%
%\bibitem{Huh:2008vj}
J.H.~Huh, J.E.~Kim and B.~Kyae,
%``Two dark matter components in N_{DM}MSSM and PAMELA data,''
arXiv:0809.2601;
% [hep-ph];
%%CITATION = ARXIV:0809.2601;%%
%\bibitem{Fairbairn:2008fb}
M.~Fairbairn and J.~Zupan,
%``Two component dark matter,''
arXiv:0810.4147;
% [hep-ph];
%%CITATION = ARXIV:0810.4147;%%
%\bibitem{Nelson:2008hj}
A.E.~Nelson and C.~Spitzer,
%``Slightly Non-Minimal Dark Matter in PAMELA and ATIC,''
arXiv:0810.5167;
% [hep-ph];
%%CITATION = ARXIV:0810.5167;%%
%\bibitem{Feldman:2008xs}
D.~Feldman, Z.~Liu and P.~Nath,
%``PAMELA Positron Excess as a Signal from the Hidden Sector,''
arXiv:0810.5762;
% [hep-ph];
%%CITATION = ARXIV:0810.5762;%%
%\bibitem{Nomura:2008ru}
Y.~Nomura and J.~Thaler,
%``Dark Matter through the Axion Portal,''
arXiv:0810.5397;
% [hep-ph];
%%CITATION = ARXIV:0810.5397;%%
%\bibitem{Ishiwata:2008cv}
K.~Ishiwata, S.~Matsumoto and T.~Moroi,
%``Cosmic-Ray Positron from Superparticle Dark Matter and the PAMELA
%Anomaly,''
arXiv:0811.0250;
% [hep-ph];
%%CITATION = ARXIV:0811.0250;%%
%\bibitem{Fox:2008kb}
P.J.~Fox and E.~Poppitz,
%``Leptophilic Dark Matter,''
arXiv:0811.0399;
% [hep-ph];
%%CITATION = ARXIV:0811.0399;%%
%\bibitem{Harnik:2008uu}
R.~Harnik and G.D.~Kribs,
%``An Effective Theory of Dirac Dark Matter,''
arXiv:0810.5557;
% [hep-ph];
%%CITATION = ARXIV:0810.5557;%%
%\bibitem{Chen:2008md}
C.R.~Chen, F.~Takahashi and T.T.~Yanagida,
%``High-energy Cosmic-Ray Positrons from Hidden-Gauge-Boson Dark Matter,''
arXiv:0811.0477;
  % [hep-ph].
  %%CITATION = ARXIV:0811.0477;%%
%\bibitem{Baek:2008nz}
S.~Baek and P.~Ko,
%``Phenomenology of $U(1)_{L_\mu - L_\tau}$ charged dark matter at PAMELA and
%colliders,''
arXiv:0811.1646;
  % [hep-ph].
  %%CITATION = ARXIV:0811.1646;%%
%\bibitem{Bi:2009md}
  X.J.~Bi, P.H.~Gu, T.~Li and X.~Zhang,
  %``ATIC and PAMELA Results on Cosmic e^\pm Excesses and Neutrino Masses,''
  arXiv:0901.0176; % [hep-ph];
  %%CITATION = ARXIV:0901.0176;%%
%\bibitem{Gogoladze:2009kv}
  I.~Gogoladze, R.~Khalid, Q.~Shafi and H.~Yuksel,
  %``CMSSM Spectroscopy in light of PAMELA and ATIC,''
  arXiv:0901.0923; % [hep-ph];
  %%CITATION = ARXIV:0901.0923;%%
J.~Mardon, Y.~Nomura, D.~Stolarski and J.~Thaler,
  %``Dark Matter Signals from Cascade Annihilations,''
  arXiv:0901.2926; % [hep-ph];
  %%CITATION = ARXIV:0901.2926;%%
%\bibitem{Phalen:2009xw}
  D.J.~Phalen, A.~Pierce and N.~Weiner,
  %``Cosmic Ray Positrons from Annihilations into a New, Heavy Lepton,''
  arXiv:0901.3165; % [hep-ph];
  %%CITATION = ARXIV:0901.3165;%%
%\bibitem{Brandenberger:2009ia}
  R.~Brandenberger, Y.F.~Cai, W.~Xue and X.~Zhang,
  %``Cosmic Ray Positrons from Cosmic Strings,''
  arXiv:0901.3474; % [hep-ph];
  %%CITATION = ARXIV:0901.3474;%%
%\bibitem{Kyae:2009jt}
  B.~Kyae,
  %``PAMELA/ATIC anomaly from the meta-stable extra dark matter component and
  %the leptophilic Yukawa interaction,''
  arXiv:0902.0071; % [hep-ph];
  %%CITATION = ARXIV:0902.0071;%%
%\bibitem{Goh:2009wg}
  H.S.~Goh, L.J.~Hall and P.~Kumar,
  %``The Leptonic Higgs as a Messenger of Dark Matter,''
  arXiv:0902.0814. % [hep-ph].
  %%CITATION = ARXIV:0902.0814;%%


\bibitem{Sommerfeld}
A. Sommerfeld, Annalen der Physik, {\bf 403}, 257 (1931).


\bibitem{Weiner}
I. Cholis, G. Dobler, D.P. Finkbeiner, L. Goodenough and N. Weiner, arXiv:0811.3641.
% [astro-ph].




\bibitem{decaying}
%\bibitem{Chen:2008qs}
C.R.~Chen, M.M.~Nojiri, F.~Takahashi and T.T.~Yanagida,
%``Decaying Hidden Gauge Boson and the PAMELA and ATIC/PPB-BETS Anomalies,''
arXiv:0811.3357;
% [astro-ph];
%%CITATION = ARXIV:0811.3357;%%
%\bibitem{Hamaguchi:2008rv}
K.~Hamaguchi, E.~Nakamura, S.~Shirai and T.T.~Yanagida,
%``Decaying Dark Matter Baryons in a Composite Messenger Model,''
arXiv:0811.0737;
% [hep-ph];
%%CITATION = ARXIV:0811.0737;%%
%\bibitem{Yin:2008bs}
P.f.~Yin, Q.~Yuan, J.~Liu, J.~Zhang, X.j.~Bi and S.h.~Zhu,
%``PAMELA data and leptonically decaying dark matter,''
arXiv:0811.0176;
% [hep-ph];
%%CITATION = ARXIV:0811.0176;%%
%\bibitem{Ibarra:2008jk}
A.~Ibarra and D.~Tran,
%``Decaying Dark Matter and the PAMELA Anomaly,''
arXiv:0811.1555;
  % [hep-ph].
  %%CITATION = ARXIV:0811.1555;%%
%\bibitem{Pospelov:2008rn}
  M.~Pospelov and M.~Trott,
  %``R-parity preserving super-WIMP decays,''
  arXiv:0812.0432; % [hep-ph].
  %%CITATION = ARXIV:0812.0432;%%
%\bibitem{Chen:2009iu}
  X.~Chen,
  %``Decaying Hidden Dark Matter in Warped Compactification,''
  arXiv:0902.0008. % [hep-ph].
  %%CITATION = ARXIV:0902.0008;%%





\bibitem{ADRS}
R. Allahverdi, B. Dutta, K. Richardson-McDaniel and Y. Santoso, arXiv:0812.2196. % [hep-ph].



\bibitem{AKM}
R. Allahverdi, A. Kusenko and A. Mazumdar, JCAP {\bf 0707}, 018 (2007).



\bibitem{ADM}
R. Allahverdi, B. Dutta and A. Mazumdar, Phys. Rev. Lett. {\bf 99}, 261301 (2007).



\bibitem{mohapatra}
%\bibitem{Mohapatra:1980qe}
R.N.~Mohapatra and R.E.~Marshak,
  %``Local B-L Symmetry Of Electroweak Interactions, Majorana Neutrinos And
  %Neutron Oscillations,''
Phys.\ Rev.\ Lett.\  {\bf 44}, 1316 (1980)
[Erratum-ibid.\  {\bf 44}, 1643 (1980)].
  %%CITATION = PRLTA,44,1316;%%


\bibitem{khalil}
S.~Khalil and H.~Okada,
  %``Dark Matter in B-L Extended MSSM Models,''
arXiv:0810.4573 [hep-ph];
  %%CITATION = ARXIV:0810.4573;%%
S.~Khalil and A.~Masiero,
  %``Radiative B-L symmetry breaking in supersymmetric models,''
  Phys.\ Lett.\  B {\bf 665}, 374 (2008).





\bibitem{tev}
T.~Aaltonen {\it et al.} [CDF Collaboration],
  %``Search for new physics in high mass electron-positron events in $p \bar{p}$
  %collisions at $\sqrt{s}$ = 1.96-TeV,''
Phys.\ Rev.\ Lett.\ {\bf 99}, 171802 (2007).
[arXiv:0707.2524 [hep-ex]].
  %%CITATION = PRLTA,99,171802;%%

\bibitem{carena}
M.S.~Carena, A.~Daleo, B.A.~Dobrescu and T.M.P.~Tait,
  %``Z' gauge bosons at the Tevatron,''
Phys.\ Rev.\  D {\bf 70}, 093009 (2004)
[arXiv:hep-ph/0408098].
  %%CITATION = PHRVA,D70,093009;%%


\bibitem{BBN}
M. Kawasaki, K. Kohri and T. Moroi, Phys. Rev. D {\bf 71}, 083502 (2005).

\bibitem{BBN2}
%\bibitem{Hisano:2009rc}
  J.~Hisano, M.~Kawasaki, K.~Kohri, T.~Moroi and K.~Nakayama,
  %``Cosmic Rays from Dark Matter Annihilation and Big-Bang Nucleosynthesis,''
  arXiv:0901.3582. % [hep-ph].
  %%CITATION = ARXIV:0901.3582;%%


\bibitem{darksusy}
%\bibitem{Gondolo:2004sc}
P.~Gondolo, J.~Edsjo, P.~Ullio, L.~Bergstrom, M.~Schelke and E.A.~Baltz,
%``DarkSUSY: Computing supersymmetric dark matter properties numerically,''
JCAP {\bf 0407}, 008 (2004).
  %[arXiv:astro-ph/0406204].
  %%CITATION = JCAPA,0407,008;%%%

\bibitem{BE}
%\bibitem{Baltz:1998xv}
E.A.~Baltz and J.~Edsjo,
%``Positron Propagation and Fluxes from Neutralino Annihilation in the Halo,''
Phys.\ Rev.\  D {\bf 59}, 023511 (1999).
  %[arXiv:astro-ph/9808243].
  %%CITATION = PHRVA,D59,023511;%%


\bibitem{NFW}
%\bibitem{Navarro:1995iw}
J.F.~Navarro, C.S.~Frenk and S.D.M.~White,
%``The Structure of Cold Dark Matter Halos,''
Astrophys.\ J.\  {\bf 462}, 563 (1996).
  %[arXiv:astro-ph/9508025].
  %%CITATION = ASJOA,462,563;%%

\bibitem{Delahaye}
%\bibitem{Delahaye:2007fr}
T.~Delahaye, R.~Lineros, F.~Donato, N.~Fornengo and P.~Salati,
%``Positrons from dark matter annihilation in the galactic halo: theoretical
%uncertainties,''
Phys.\ Rev.\ D {\bf 77}, 063527 (2008).
  %[arXiv:0712.2312 [astro-ph]].
  %%CITATION = PHRVA,D77,063527;%%

\bibitem{CRgamma}
%\bibitem{Jeltema:2008vu}
  T.E.~Jeltema, J.~Kehayias and S.~Profumo,
  %``Gamma Rays from Clusters and Groups of Galaxies: Cosmic Rays versus Dark
  %Matter,''
  arXiv:0812.0597; % [astro-ph];
  %%CITATION = ARXIV:0812.0597;%%
P.~Meade, M.~Papucci and T.~Volansky,
  %``Dark Matter Sees The Light,''
  arXiv:0901.2925. % [hep-ph].
  %%CITATION = ARXIV:0901.2925;%%

\bibitem{CRneutrino}
%\bibitem{Hisano:2008ah}
  J.~Hisano, M.~Kawasaki, K.~Kohri and K.~Nakayama,
  %``Neutrino Signals from Annihilating/Decaying Dark Matter in the Light of
  %Recent Measurements of Cosmic Ray Electron/Positron Fluxes,''
  arXiv:0812.0219. % [hep-ph].
  %%CITATION = ARXIV:0812.0219;%%



\bibitem{cdms} J.~Yoo  [CDMS Collaboration],
  %``Results from the CDMS 5-Tower Operation,''
arXiv:0810.3527. % [hep-ex].
  %%CITATION = ARXIV:0810.3527;%%

\bibitem{Direct}
L. Baudis, arXiv:0711.3788; % [astro-ph];
T. Saab, talk at Dark2009 Conference, Christchurch, New Zealand.


\bibitem{endpoint}
I.~Hinchliffe, F.E.~Paige, M.D.~Shapiro, J.~Soderqvist and W.~Yao,
  %``Precision SUSY measurements at LHC,''
Phys.\ Rev.\ D {\bf 55}, 5520 (1997).
  %[arXiv:hep-ph/9610544].
  %%CITATION = PHRVA,D55,5520;%%

\bibitem{AN}
J. R. Ellis, G. Ridolfi and F. Zwirner, Phys. Lett. B {\bf 257}, 83 (1991); R. Arnowitt and P. Nath, Phys. Rev. D {\bf 46}, 3981 (1992).



\bibitem{ECAL}
T.G.~Guzik, {\it et al.},
% The Electron Calorimeter (ECAL) Long Duration Ballon Experiment,
in ICRC 2007 Proceedings.


\bibitem{atmos}
Y. Fukuda {\it et al.} [Super-Kamiokande Collaboration], Phys. Rev. Lett. {\bf 81}, 1562 (1998), hep-ex/9807003.


\end{thebibliography}
\end{document}